\documentclass[twocolumn,showpacs,amsmath,prl]{revtex4}

\usepackage{graphicx}

\begin{document}
\title{Percolation, depinning, and avalanches in capillary
condensation of gases in disordered porous solids}

\author{M. L. Rosinberg, E. Kierlik, and G. Tarjus}
\affiliation{Laboratoire de Physique  Th{\'e}orique des Liquides,  Universit{\'e} Pierre et
Marie Curie, 4 place Jussieu, 75252 Paris Cedex 05, France}

\begin{abstract}
We propose a comprehensive theoretical description of hysteresis in
capillary condensation of gases in mesoporous disordered
materials. Applying mean-field density functional theory to a
coarse-grained lattice-gas model, we show that the morphology of
the hysteresis loops is influenced by out-of-equilibrium transitions
that are different on filling and on draining. In particular,
desorption may be associated to a depinning process and be 
percolation-like without explicit pore-blocking effects.

\end{abstract}
\maketitle

\def\be{\begin{equation}}
\def\ee{\end{equation}}
\def\bea{\begin{eqnarray}}
\def\eea{\end{eqnarray}}

Hysteresis in the capillary
condensation of gases adsorbed in disordered mesoporous materials has been
widely studied, but its theoretical interpretation remains 
controversial.
Experimentally, it is mainly observed in sorption isotherms measuring 
the amount of fluid present  in the 
solid as the pressure of the ambient vapor (or the
chemical potential $\mu$) is gradually increased and then
decreased. Experiments have been performed  in various adsorbents, from
low-porosity  glasses  to very light silica aerogels,  and the main
observations can be summarized as follows: i) the hysteresis loop shrinks as the
temperature increases and vanishes above a certain  temperature
$T_h$\cite{E1967,ball89,G1999}; ii) draining occurs over a narrow
range of pressures, which usually results in an asymmetric shape of the loop;
iii) the  shape also depends on the geometric and energetic properties
of the solid
and may change from smooth to rectangular as the porosity
increases\cite{tul99}; iv) when the hysteresis is smooth, there is a whole family of scanning
curves obtained by performing incomplete filling-draining
cycles\cite{E1967,lilly93}, and in the early stages of the drainage, the
emptied  regions exhibit long-range fractal correlations, reminiscent of invasion percolation\cite{weitz93}. 
The conjunction of these observations cannot be explained by the various
mechanisms of hysteresis that have been proposed in the literature,
either modelling  the solid as a collection of  independent
pores\cite{E1967} or focusing on kinetic network effects  (e.g., pore-blocking)\cite{M1983}. In an
effort to provide a single unifying theoretical framework, we have
recently presented a new description of capillary condensation in disordered
materials that allows for the study of both equilibrium and
out-of-equilibrium phenomena\cite{K2001}.  Applying mean-field density functional
theory (i.e., local mean-field theory or MFT) to a lattice-gas model that
incorporates at  a coarse-grained  level the  geometric  and energetic
disorder of an interconnected porous structure, we have shown that  the experimentally observed hysteretic
behavior is related to the appearance  of  a complex
free-energy landscape with a  large number  of  metastable
states; the main result is that
hysteresis can occur both with and without a true equilibrium phase transition. 

In  this letter, we use the same theoretical framework to show that
the morphology  of the hysteresis loop is  affected
by {\it out-of-equilibrium} phase  transitions (which differ on
adsorption and desorption) and that the 
presence of the interface between  the gas reservoir and the porous
solid has a dramatic influence on the hysteretic behavior.  A major
finding of our study is that desorption can involve a percolation process without  invoking  additional  kinetic
mechanisms such as  pore blocking.  To understand the changes of
morphology  of the  sorption  isotherms  and  the  out-of-equilibrium
collective phenomena, we   borrow  from studies   on   avalanches and
criticality in   low-T     ferromagnetic materials ({\it i.e.}, Barkhausen
noise)\cite{sethna93,planes99,tadic96} and on driven interfaces  and depinning   in  systems   with quenched
disorder\cite{robbins92,nowak98}.    Whereas  most of   these
studies  are at  T=0, MFT allows  us to  consider   finite
temperatures, which are  relevant  for capillary condensation,  and to
relate  such phenomena as avalanches  and  interface depinning to  the
properties of the underlying free-energy surface.

\begin{figure}
\begin{center}
\resizebox{7cm}{!}{\includegraphics{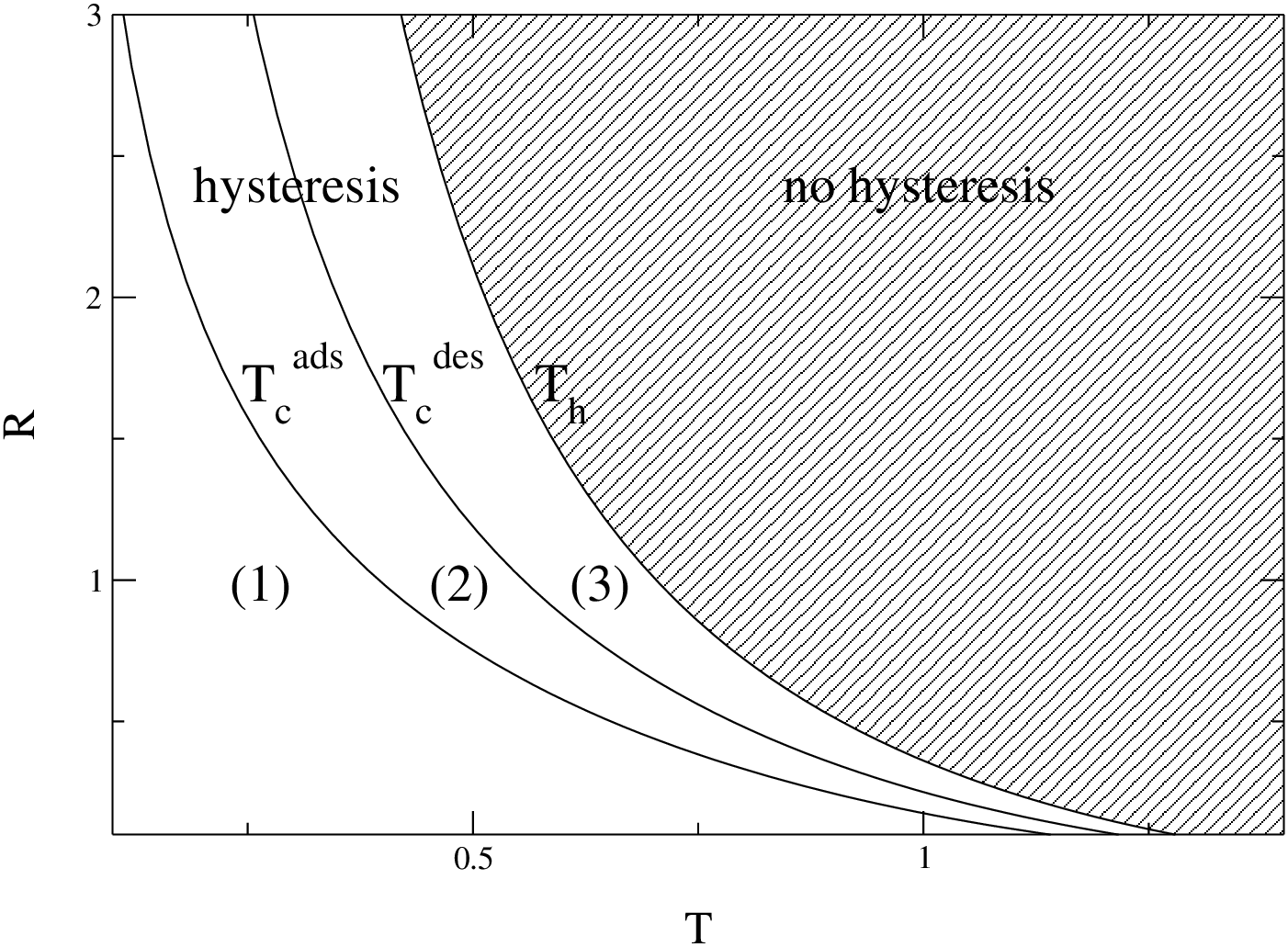}}
\caption{Schematic out-of-equilibrium phase diagram of the model
showing the different regimes discussed in the text. $R$ parametrizes
the strength of the solid-induced perturbation (the units are arbitrary).}

\end{center}
\end{figure}
The hysteresis  behavior  seen experimentally can  be rationalized by
considering the phase diagram   plotted in Fig. 1, that  schematically
summarizes  the  present study. More details will be given in a
forthcoming publication\cite{DKRT2002}. $R$ parametrizes 
the effective strength of the solid-induced perturbation
and  may depend in a complicated fashion on the porosity and
on the solid-fluid interaction. Within MFT, there are lines
of  out-of-equilibrium critical points separating  a region (1) (low
$T$, low $R$)  in which both the adsorption and desorption isotherms
have a jump that represents  a first-order-like transition (an
infinite avalanche), an  intermediate region (2) in which the adsorption
is smooth and the desorption discontinuous, and a region (3) (high
$T$, high $R$) in which the two branches are smooth and the desorption  is
essentially percolation-like. These collective
phenomena take place below  $T_h$ at which hysteresis  first appears.
(The   equilibrium   transition   discussed previously\cite{K2001}  is   not   considered here.)
These sharp out-of-equilibrium  transitions are well defined  in MFT
 in which thermally activated processes  are neglected.  In
experiments, these phenomena are observable when the experimental time
scale is  smaller than the  equilibration  time.  The fact that the measured
hysteresis loops are quite reproducible\cite{E1967,ball89,G1999} and closely resemble
those predicted by MFT  strongly supports the  existence of such   a
separation  of  time scales. Note also that  avalanche
events associated to capillary condensation are indeed observed\cite{lilly93,lilly96}.  

As in previous studies\cite{K2001}, we use a disordered lattice-gas model  whose Hamiltonian is
\bea
{\cal H} = &-&w_{ff}\sum_{<ij>} \tau_{i}\tau_{j} \eta_i \eta_j \nonumber\\
&-&w_{mf}\sum_{<ij>} [\tau_{i}\eta_i (1-\eta_j)+\tau_{j}\eta_j (1-\eta_i)] 
\eea	
where  $\tau_i=0,1$ and $(1-\eta_i)=0,1$ denote the fluid and matrix occupation
variables, respectively,  and $<ij>$ stands for nearest-neighbor pairs of sites. 
We  consider the simplest matrix  microstructure in which matrix ``particles''
are distributed randomly over a 3-dimensional lattice of $N$ sites. The  influence of  the matrix is then specified
by two parameters, the average  matrix density $\rho_m$ that controls the
porosity, and the interaction ratio $y=w_{mf}/w_{ff}$ that controls
the wetting properties of the solid/fluid  interface. The results are
qualitatively similar for more realistic matrix microstructures\cite{DKRT2002}.
The lattice is periodically replicated in all
directions. To study surface effects we consider two
different setups, by letting or not a slab of width
$\Delta L_x$ free of matrix sites: this defines a region (a ``reservoir'') that only contains bulk vapor (at least
for  the thermodynamic conditions under study) with two planar
interfaces  in the x-direction (see Fig. 3). 

For a given realization $\{\eta_i\}$ of the random matrix, the fluid
density profile $\{\rho_i\}$ is obtained from minimization of the
mean-field grand-potential functional\cite{K2001}, which yields a set
of $N$ coupled equations, 
\be
\rho_i=\eta_i[1+e^{-\beta (\mu+w_{ff}\sum_{j/i}[\rho_j+y(1-\eta_j)]) }]^{-1} \ ,
\ee
where $\beta=1/(k_BT)$ and the sum is over the nearest-neighbors of site
$i$. These nonlinear equations are solved by a simple iterative
method, increasing  or decreasing progressively $\mu$  to
obtain the sorption  isotherms ({\it i.e.}, $ \rho_f( \mu,T)=(1/N) \sum_i\rho_i$) or performing  more complicated
trajectories to obtain the scanning curves. The results are illustrated
here for a bcc lattice, either for a single sample of 
linear size $L=96$  ($N=2L^3$) with $\Delta L_x=12$, or, when the
slope of the sorption isotherms is steep and  exhibit strong variations
with $L$,  through a finite-size scaling analysis 
after averaging over several hundreds of matrix realizations.  

\begin{figure}
\begin{center}
\resizebox{8cm}{!}{\includegraphics{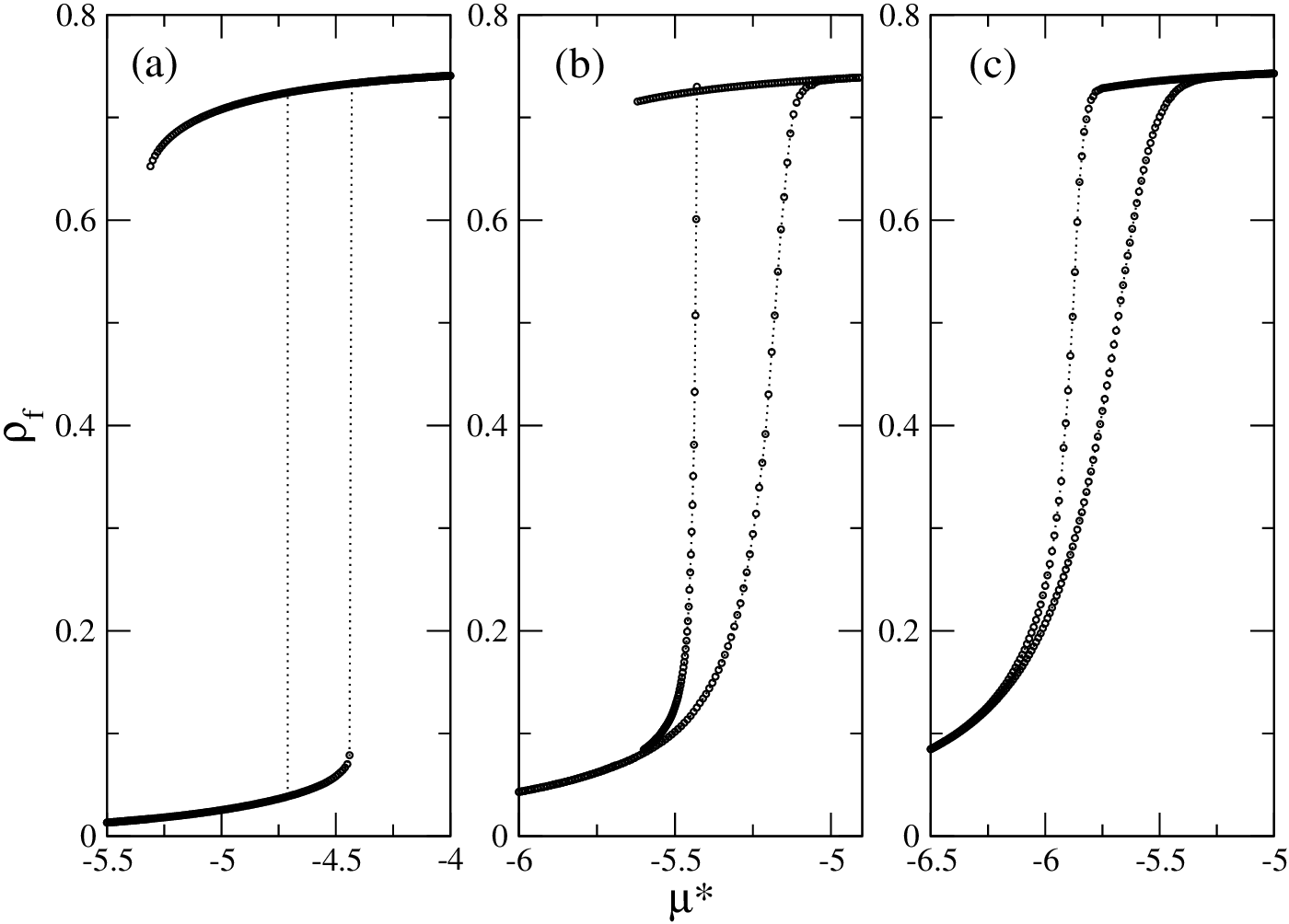}}
\caption{Hysteresis loops in absence or presence of an external
interface ($\mu^*=\mu/w_{ff}$): (a) $y=0.8$, (b) $y=1.2$, (c)
$y=1.5$. Note in Figs. a and b the artificial extension of the
desorption branch in the absence of interface.}

\end{center}
\end{figure}

The shape of the hysteresis loop depends on $T,  \rho_m$ and $y$. Here,
for the sake of clarity,  we only
discuss  the variation with the interaction
ratio $y$ while keeping $T$ and $\rho_m$ constant (then $R\equiv y-1/2$  controls the strength of the
local random fields in the system\cite{K2001}).  
Typical results obtained for $T^*=k_BT/w_{ff}=0.8$
and $\rho_m=0.25$ are shown in  Fig. 2. One first notices that introducing an
external interface has a major influence on the desorption branch, the
adsorption branch being itself unaffected (see also\cite{SM2002}).
This influence is strongest at small $y$  and vanishes for $y= 1.5$: the presence of
the  surface reduces the  range  of stability  of the  metastable
adsorbed  liquid  (the flat,  upper  part of  the desorption
curve in Figs. 2a,b),
thereby narrowing the hysteresis loop. One can also see that
 the discontinuous jumps present in the
sorption isotherms for $y=0.8$ (region (1) of Fig. 1) are replaced by 
smooth variations for $y=1.2$ and $1.5$ (region (3) of Fig. 1) and that  the
hysteresis loops become  asymmetric, as observed in experiments. 
\begin{figure}
\begin{center}
\resizebox{8cm}{!}{\includegraphics{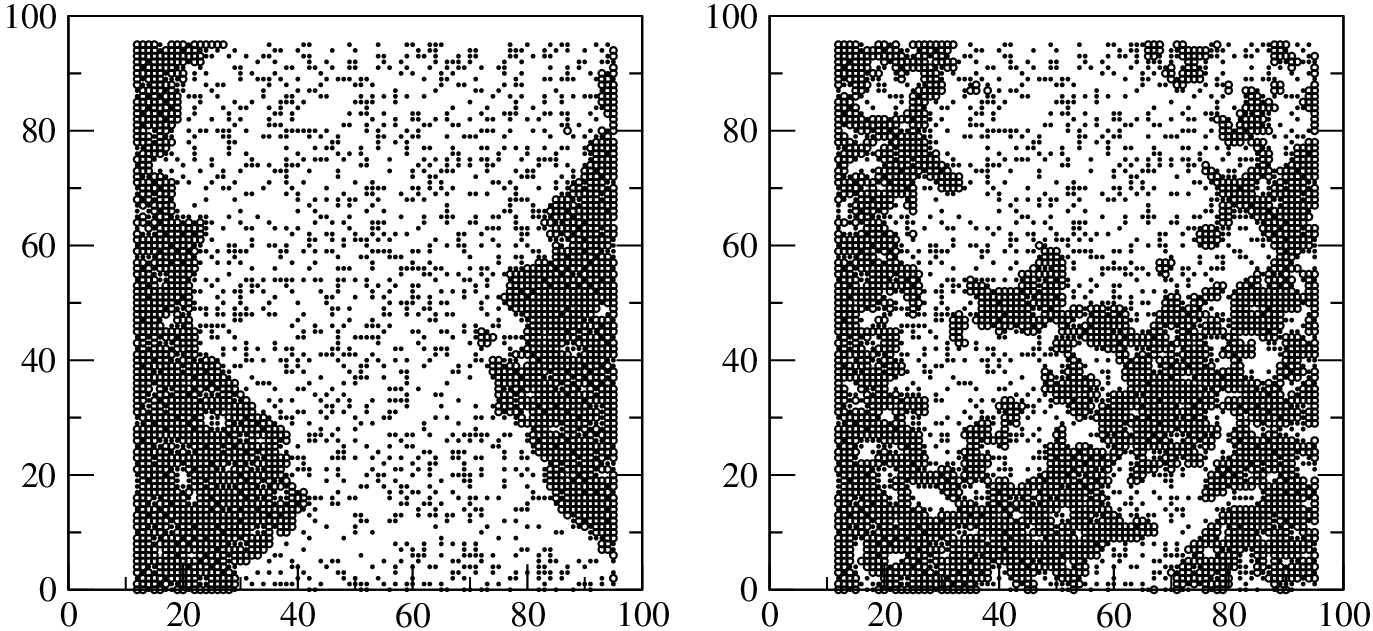}}
\caption{Cross-section of a 3-D system of linear size $L=96$ during
desorption right at the depinning transition: (a) $y=0.8, \mu^*=-4.705$, (b)
$y=1.2, \mu^*=-5.44$. The initial interfaces with the bulk vapor
are located at $x=12$ and $x=96$. The black regions represent
the drained  sites ($\rho_i<0.3$) and the grey dots the matrix sites.}
\end{center}
\end{figure}
 
Why is the desorption so  sensitive to the presence of the external surface? 
For pores  of simple geometry, one knows  that draining is facilitated
by the contact with the gas reservoir because of the presence of a
meniscus that, at  a given $\mu$, can traverse
the whole pore at no free-energy cost. However, this picture cannot be
simply transposed to disordered  porous materials: this is illustrated
by the  large-$y$ case  for which  the  interface has  no  effect  at
all. The key-concept for understanding the results  shown in Fig. 2 is
that of pinning and depinning of a driven interface in the presence of
quenched disorder. Consider the desorption process when slowly
decreasing $\mu$ from $\mu_{sat}$, the chemical potential at bulk liquid-gas coexistence.  At
first, the interface between the external vapor and the adsorbed fluid is pinned by the disorder, {\it
i.e.},  it only enters superficially in the material and this latter
stays macroscopically filled with liquid. As one further reduces $\mu$, several  situations are encountered. For a
large enough  interaction ratio $y$, bubbles  of gas may appear in the
bulk of the porous medium, and the mass adsorbed 
decreases continuously as these bubbles grow, coalesce, and eventually
extend over the whole  pore space; this phenomenon is insensitive to the presence
of the surface, as illustrated by Fig. 2c.   For  smaller values  of $y$,  no
bubbles of gas  appear in the  bulk of the  material. This can be
inferred from the fact that the same system without external surface 
does not drain until much smaller  values of $\mu$ are reached (Figs.  2a,b).
The  draining mechanism is thus the depinning of the 
interface with the bulk vapor: for a certain chemical potential
$\mu_c$,  the  driving  force on the  interface is large
enough to erase the free-energy barriers,  and the interface  sweeps through the
whole system, which results in the rapid drop seen in the desorption isotherm.

\begin{figure}
\begin{center}
\resizebox{7cm}{!}{\includegraphics{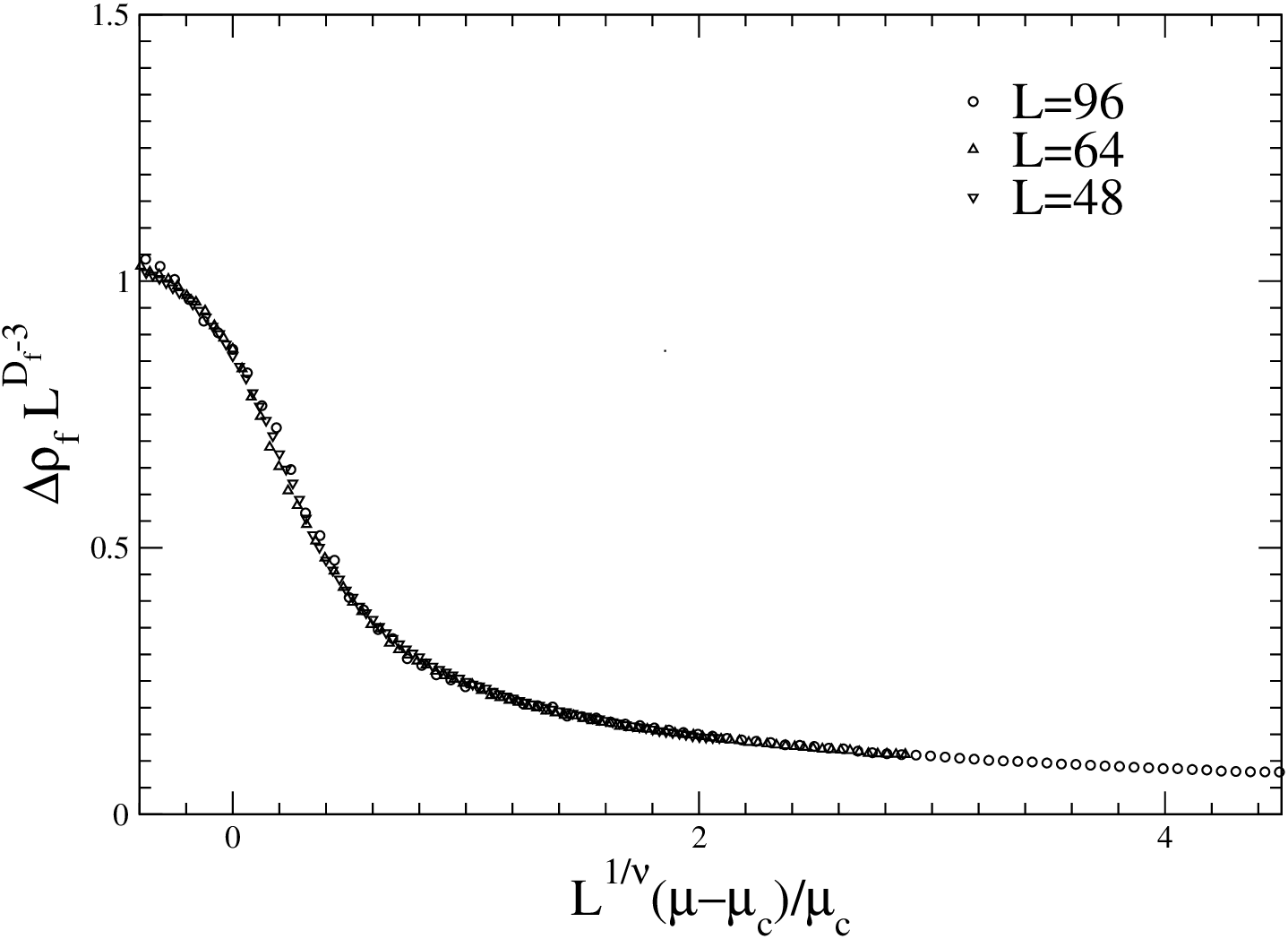}}
\caption{Finite-size-scaling collapse of $\Delta \rho_f$ on desorption for
$y=1.2$ with $\mu_c^*=-5.446, \nu=0.89, D_f=2.84$ ($\Delta L_x=L/8$).}

\end{center}
\end{figure}
Two different  cases can  occur as  $\mu$ approaches $\mu_c$ from
above. For a small  interaction  ratio (Fig. 2a), the  surface of the
growing gas region inside the solid keeps a well-defined orientation at long
length  scales (that of  the initial  interface with the bulk vapor),
and, exactly at $\mu=\mu_c$, the desorption  proceeds
discontinuously, as shown in Fig. 2a\cite{Rem}. For intermediate
values of $y$ (Fig. 2b), the growing  interface has no well-defined orientation;
it is rather isotropic and  self-similar, and the desorption is rapid (at least
in the early stage of draining), but  continuous, as  in invasion   percolation.   The former  case
corresponds to a  compact growth of the gas region with a self-affine
interface  and the  latter  to a self-similar    growth, characteristic  of percolation: this
is illustrated  in Fig.  3 showing the drained regions where $\rho_i<0.3$
(since for a large majority of sites, $\rho_i$ is found either
lower than $0.3$ or larger than $0.8$, drained sites can be
conveniently defined by this threshold\cite{DKRT2002}). It should be stressed that the
percolation-dominated  draining process which is observed for 
intermediate  values  of  $y$ does not require the introduction of
kinetic constraints  such as pore blocking.  The  experimental
observations for hexane  in Vycor\cite{weitz93} can thus be interpreted
without invoking this latter mechanism. 

One can  extract the 
 exponents controlling the critical behavior as $\mu\to \mu_c$ by
performing a finite-size scaling analysis of 
$\Delta\rho_f(\mu)=\rho_f(\mu_{sat}) -\rho_f(\mu)$. This is shown in Fig.  4 for the
percolation regime: there, the value of $\nu$ is
compatible with that of 3D percolation, whereas the value of the
fractal dimension
$D_f$ is somewhat larger, for reasons as yet unclear. A similar
analysis can be carried out in the self-affine regime, and
more details will be given elsewhere\cite{DKRT2002}. 
At  a critical value of $y$ somewhere between $0.8$ and $1.2$, there is a morphology transition from
self-affine to self-similar growth, which  corresponds to a transition
from  a  discontinuous desorption to a continuous, percolation-like desorption.   This is similar to what is 
observed for domain growth  in random magnets or in  fluid invasion\cite{robbins92}, $y$ playing the  role
of the parameter  controlling the wetting properties of the invading liquid. 

Since the filling process is not affected by the presence of  the
 interface with the gas reservoir, the jump in the adsorption isotherm observed in
Fig. 2a has not the same origin as that in desorption (note that we had to perform  a finite-size scaling study to
conclude on the existence of a jump  in the thermodynamic
limit\cite{DKRT2002}). It corresponds to the appearance of a
macroscopic, connected network of liquid regions in the bulk of the
porous medium. In MFT, this is again a  {\it  bona  fide} phase
transition,  albeit out-of-equilibrium: it is  similar to the infinite
avalanche  seen in   the   Barkhausen noise  of  low-T  ferromagnetic
materials and discussed in Refs.\cite{sethna93,planes99,tadic96}. One then  expects that the
adsorption isotherm should change from discontinuous to continuous  at
some critical value of $y$ (see Fig. 2).

Similar modifications  of the hysteresis loop are observed when
changing $T$ at constant $y$  and $\rho_m$\cite{DKRT2002}: at low T there
are  macroscopic avalanches on draining and on filling that disappear at two different  critical
temperatures, $T_c^{ads}(R)$ and $T_c^{des}(R)$; for higher $T$,
both the adsorption and the desorption branches are continuous, and at
$T_h(R)$ the two branches merge and the isotherm become fully
reversible. This leads to the phase diagram illustrated in Fig. 1. Similar changes  in the shape of the loop also
occur when varying the porosity at constant $y$\cite{DKRT2002}. This may account
for the experimental observations on aerogels\cite{tul99}.

\begin{figure}
\begin{center}
\resizebox{4.5cm}{!}{\includegraphics{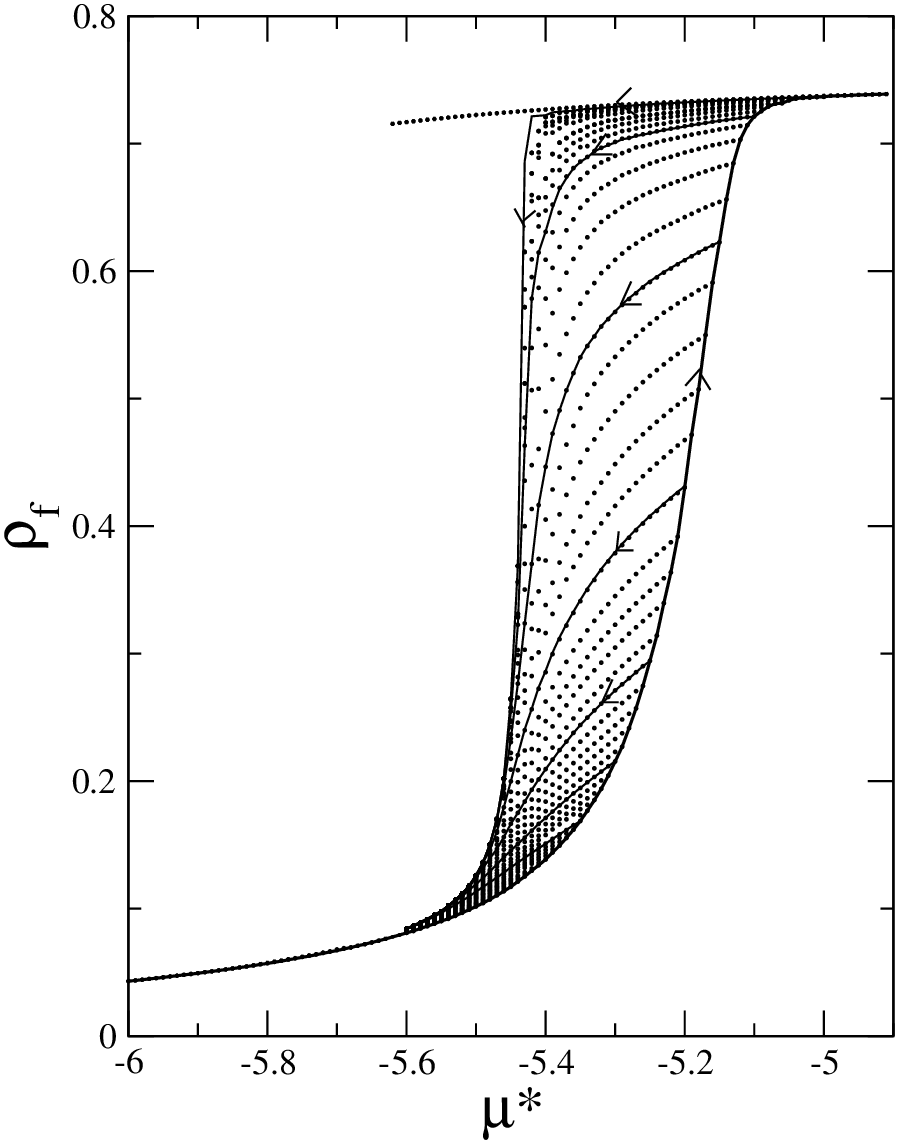}}
\caption{Metastable states obtained by performing desorption scanning
trajectories in the absence of an external interface for
$y=1.2$. Compare with Fig. 2b.}

\end{center}
\end{figure}
Finally, we have also studied the connection between the  above results and the
characteristics  of the free-energy   landscape. Below the hysteresis
line $T_h(R)$ (see  Fig. 1), MFT predicts  the  appearance of a
large number of  metastable  states  (minima of  the   grand-potential
functional)   in  which   the   system can  get
trapped. Untrapping only occurs under the influence of the external driving
force, {\it i.e.}, the change in $\mu$.  We  again emphasize that this
picture is valid  provided the thermally activated   processes  for
relaxation (processes that are for instance responsible for the creep
motion of pinned interfaces\cite{nowak98}) are slower than the experimental timescale.  The metastable
states  are responsible for the  hysteresis loop discussed above
as well  as  for the  scanning curves and the inner subloops. Fig. 5
displays some of these 
states obtained by performing desorption scanning trajectories in the
system without surface for $y=1.2$ (additional  metastable states could be
obtained by the procedure explained in Ref.\cite{K2001}). It is
remarkable that the 
hysteresis loop  obtained  in the presence of the external surface appears
as the true envelope of the scanning curves (compare with Fig. 2b), an
observation that holds for all the cases studied.

In summary, by combining mean-field density functional theory and a
coarse-grained lattice-gas model, we have provided a comprehensive
picture of hysteresis in capillary condensation of gases in disordered
mesoporous solids. We have shown that the change of morphology of the
hystereresis loop is related to the occurence of out-of-equilibrium
phase transitions whose nature is different on adsorption and on
desorption: this is due to the presence of an interface between the porous
matrix and the gas reservoir.

\quad
The Laboratoire de Physique Th{\'e}orique des Liquides is the UMR 7600 of
the CNRS.


\begin{references}
\bibitem{Rem} Note that the depinning transition is a critical
transition\cite{robbins92,nowak98}: the average distance traveled by
the interface before pinning occurs diverges as a power law when $\mu\to \mu_c$.
\end{references}
\end{document}